# Comparative Analysis of Black-Box and White-Box Machine Learning Model in Phishing Detection

Abdullah Fajar [1)*] , Setiadi Yazid *[2)] , Indra Budi[3)]

[1)2)3)]*Universiti Indonesia, Depok, Indonesia*

[*1)]abdullah.fajar@ui.ac.id, [2)]setiadi@cs.ui.acid, [3)]indra@cs.ui.acid

[1)] *Universitas Telkom, Bandung, Indonesia*

[1)]abdfajar@telkomuniversity.ac.id

*Abstract*

**Background:** Explainability in phishing detection model can support a further solution of phishing attack mitigation by increasing trust and understanding how phishing can be detected.
**Objective:** The aims of this study to determine and best recommendation to apply an approach which has several components with abilities to fulfil the critical needs
**Methods:** A methodology starting with analyzing both black-box and white-box models to get the pros and cons specifically in phishing detection. The conclusion of the analysis will be validated by experiment using a set of well-known algorithms and public phishing datasets. Experimental metrics covers 3 measurements such as predictive accuracy and explainability metrics.
**Conclusion:** Both models are comparable in terms of interpretability and consistency, with room for improvement in diverse datasets. EBM as an example of white-box model is generally better suited for applications requiring explainability and actionable insights. Finally, each model, white-box and black-box model has positive and negative aspects both for performance metric and for explainable metric. It is important to consider the objective of model usage.

## I. INTRODUCTION

The detection of phishing attacks has become a critical task in the field of cybersecurity, as these deceptive tactics continue to evolve and pose significant risks to individuals and organizations [1]. Machine learning models have emerged as a promising approach to address this challenge, with researchers exploring various techniques to improve the accuracy and reliability of phishing detection[2].

One key aspect of this endeavor is the ability to generate explanations for the model's predictions, which is particularly important in safety-critical domains like cybersecurity. Explainable artificial intelligence has gained attention as a means to address the "black box" nature of many machine learning models, where the internal decision-making process is not readily interpretable.[3], [4]

Recent research has explored the trade-offs between black-box and white-box machine learning models in the context of phishing detection[5], [6]. The ability to generate explanations for the model's predictions is particularly important in safety-critical domains like cybersecurity, as it can enhance trust and accountability in the detection system[7]. Explainable artificial intelligence has gained attention to address the "black box" nature of many machine learning models, where the internal decision-making process is not readily interpretable. By generating explanations, the factors driving the model's decisions can be better understood, which is crucial for understanding and validating the model's behavior, especially in high-stakes applications like phishing detection.[8]

In previous works, several research has questioned related to the reason why need generating explanations in phishing detection, here are the questions:
1. How to Explain Phishing Attempts? A crucial question in the field is: "How is this website attempting to steal the identity of a well-known brand?" [9]. This question highlights the need for systems that can not only detect phishing attempts but also provide clear explanations of the methods used by attackers.

---

[*] Corresponding author

2. How to Enhance User Understanding? Researchers are exploring ways to improve users' comprehension of phishing threats. The question of how to generate coherent and complete explanations in natural language for anti-phishing system decisions is a key focus [10]. This aims to address the issue of users ignoring warnings due to lack of significant information.
3. How to Identify Crucial Phishing Information? An important question is how to automatically learn and identify the most important and phishing-relevant information in suspicious data [11]. This is particularly relevant for email phishing detection, where pinpointing specific sentences or elements that indicate phishing is crucial.
4. How to Provide Actionable Insights? Researchers are investigating how to offer insights into algorithm predictions, enabling humans to understand and evaluate why a particular URL is not secure to visit [12]. This question relates to the broader field of Explainable AI (XAI) in cybersecurity.
5. How to Select Relevant Features? A significant question is how to select the most important features for phishing detection models while ensuring they remain interpretable [13]. This involves balancing the need for accuracy with the requirement for explainability.
6. How to Design Effective Warning Dialogs? Researchers are exploring how to create warning dialogs that not only alert users about possible attacks but also explain why a website is suspicious [14]. This question addresses the limitations of existing warning systems and aims to improve user defense against phishing.
7. How to Build Trust in Detection Systems? A critical question is how to increase user trust in phishing detection systems through explanations [15]. This involves finding ways to make AI-based solutions more transparent and understandable to users, potentially increasing the effectiveness of phishing prevention.

Summarizing the answers of those questions, AI-based phishing detection systems have proven highly effective in combating phishing attacks, leveraging advanced techniques to identify and predict vulnerabilities. The effectiveness of these systems is further enhanced by incorporating explainability features, which provide users with insights into the reasoning behind phishing warnings. This transparency not only improves user understanding but also fosters trust in the detection system.

Furthermore, research highlights the importance of clear and concise explanations that resonate with users, enabling them to make informed decisions about potential threats. The development of robust and interpretable models, coupled with user-centric design principles, is crucial for building reliable and trustworthy anti-phishing solutions. This highlights a critical need for phishing detection systems to not only be accurate but also to clearly and understandably explain their decisions to users. This will improve user comprehension, trust, and ultimately, their ability to stay safe online.

In this paper, the aims to determine and best recommendation to apply an approach which has several components with abilities to fulfil the critical needs. Achieving the goals, this paper proposed a methodology starting with analyzing both black-box and white-box models to get the pros and cons specifically in phishing detection. The conclusion of the analysis will be validated by experiment using a set of well-known algorithms and public phishing datasets. Experimental metrics covers 3 measurements such as predictive accuracy and explainability metrics.

The rest of this paper covers result description that explain result of comparison and findings in experimental validation. The closing part is described by discussion and conclusion.
.

## II. Literature Review

### A. Phishing Attack Landscape and General Detection Approaches

#### 1) Phishing Attacks Landscape

Phishing attacks have become increasingly sophisticated in recent years, employing a diverse arsenal of techniques to deceive victims and gain unauthorized access to sensitive information or systems. These attacks manifest in various forms, including email-based phishing, which typically involve fraudulent messages impersonating legitimate entities; SMS-based phishing, where attackers leverage text messages to lure victims; and social media-based phishing, where threat actors create fake accounts or pages to exploit users' trust. Additionally, phishers have adapted to new communication channels, such as instant messaging platforms and voice calls, further expanding the threat landscape. The growing prevalence of these attacks, combined with their evolving tactics, underscores the critical need for robust phishing detection and prevention measures to safeguard individuals and organizations against these persistent cyber threats.[16], [17]

While phishing attacks have indeed become more sophisticated in recent years, employing a diverse array of techniques to deceive victims, there is also a strong argument to be made that our defenses against these threats have been steadily improving. Advances in machine learning, data analytics, and cybersecurity best practices have enabled organizations to develop increasingly effective phishing detection and prevention measures. Robust email filters, behavioral monitoring systems, and user education programs have all contributed to a stronger security posture, helping to safeguard individuals and organizations against these persistent cyber threats. Additionally, the growing prevalence of these attacks may be offset by the increased awareness and vigilance of both users and security professionals, as they work to stay ahead of the evolving tactics of phishers[18], [19], [20]. Ultimately, the fight against phishing is an ongoing battle, but one in which the defensive capabilities of modern cybersecurity solutions are steadily advancing.

*2) General Phishing Detection Approaches*

Numerous approaches have been explored to detect and mitigate phishing attacks, including rule-based systems, heuristic-based techniques, and machine learning-based methods. Rule-based systems rely on predefined sets of rules or signatures to identify known phishing patterns, but they often struggle to keep pace with the rapid evolution of phishing tactics. These systems typically have a limited ability to adapt to new types of phishing attacks, as they are primarily designed to detect known attack patterns[19], [21], [22]. Heuristic-based approaches, on the other hand, leverage a combination of content-based and behavioral analysis to identify suspicious characteristics of potential phishing attempts. These techniques examine various features, such as the URL structure, the presence of misleading or suspicious content, and user interaction patterns, to assess the likelihood of an attack. While heuristic-based methods can be more flexible than rule-based systems, they may still have difficulty keeping up with the constantly changing tactics of phishers.

The increasing sophistication of phishing attacks has driven a growing interest in machine learning-based solutions, which have the potential to adapt to new threats and detect previously unseen patterns. These models, trained on large datasets of phishing and legitimate samples, can learn to distinguish between legitimate and malicious content, often outperforming traditional rule-based and heuristic-based methods in terms of accuracy and adaptability. Machine learning-based phishing detection approaches can leverage advanced techniques, such as deep learning and ensemble methods, to capture complex patterns and relationships within the data. This can lead to enhanced detection capabilities, allowing the models to identify and flag novel phishing threats more effectively compared to rule-based or heuristic-based systems[23], [24], [25], [26]. Additionally, machine learning models can continuously learn and adapt as new phishing techniques emerge, improving their performance over time and maintaining their effectiveness against the evolving threat landscape.

Furthermore, the use of interpretable machine learning models can provide valuable insights into the decision-making process of the phishing detection system. By generating explanations for the model's predictions, the factors driving the detection can be better understood, which is crucial for validating the system's behavior and fostering user trust, especially in high-stakes applications like cybersecurity[27], [28]. Explainable AI techniques, such as feature importance analysis, local interpretability methods, and model-agnostic explanations, can shed light on the key factors the model considers when determining if a given input is a phishing attempt or not. This transparency can help security teams and end-users comprehend the reasoning behind the system's decisions, enabling them to better assess the reliability and trustworthiness of the phishing detection model[27], [29].

Moreover, the deployment of interpretable and trustworthy machine learning models for phishing detection requires a comprehensive approach that considers the practical implications and user needs[2]. Developers should focus on iterative model refinement, providing contextual explanations, integrating the system into existing workflows, fostering collaborative development, and offering ongoing training and support. By adopting this holistic approach, organizations can successfully implement interpretable phishing detection solutions that enhance security, increase user confidence, and maintain the effectiveness of the system against the evolving threat landscape[30].

*B. Explainability in the Real-World Use of Machine Learning*

There are many reasons why explainability is crucial in real-world machine learning practice. It's key for fostering trust, guaranteeing fairness, enabling model debugging and improvement, and achieving regulatory compliance. Here's a closer look:

1. **Trust and Transparency:** Users are more likely to trust and accept decisions made by a model if they understand how those decisions were reached. Explainability helps build this trust by providing insights into the model's reasoning[21].

2. **Fairness and Bias Detection:** Machine learning models can inherit biases present in the training data, leading to unfair or discriminatory outcomes. Explainability allows us to identify and mitigate these biases by understanding which features the model relies on[22]. [

3. **Debugging and Improvement:** When a model makes a mistake, explainability helps us understand why. This understanding is essential for debugging the model, identifying areas for improvement, and building more robust and reliable systems[23].

4. **Regulatory Compliance:** In certain domains, such as healthcare and finance, regulations require transparency and accountability in decision-making processes. Explainable AI ensures compliance by providing auditable explanations for model predictions[23].

5. **User Acceptance and Adoption:** Explainable AI fosters user confidence and encourages the adoption of machine learning systems. When users understand how a model works, they are more likely to use it effectively and trust its recommendations[21].

In essence, explainability bridges the gap between complex machine learning models and human understanding. It promotes responsible AI development and deployment, ensuring that these powerful technologies are used ethically and effectively.

*1) Explainability White-Box Machine Learning Models*

Explainability in a white-box machine learning model refers to the inherent transparency of its decision-making process. Unlike black-box models where the internal workings remain hidden, white-box models, also known as interpretable models, allow us to understand how they arrive at their predictions.

Here's a breakdown of the explainability aspects:

1. **Model Structure:** White-box models typically have a simple and understandable structure, such as linear regression, decision trees, or rule-based systems. This inherent simplicity makes it easier to trace the path from input features to the final prediction[24].

2. **Feature Importance:** These models can readily provide insights into which features are most influential in driving the predictions. For instance, in a decision tree, the features at the top nodes have a higher impact on the final decision[25], [26].

3. **Rule Extraction:** Some white-box models, like decision trees or rule-based systems, explicitly represent their decision logic as a set of rules. These rules can be easily understood by humans, providing a clear explanation for each prediction[27].

4. **Local Interpretability:** We can often obtain explanations for individual predictions. For example, in a linear regression model, we can see how each feature contributes to the final prediction for a specific data point[6].

5. **Global Interpretability:** White-box models allow for an overall understanding of how the model behaves across the entire dataset. This global perspective helps identify potential biases or limitations in the model's decision-making[28].

It's important to note, however, that there are trade-offs to consider with white-box models. Simpler, more interpretable models may not always achieve the same level of predictive accuracy as complex black-box models, particularly when dealing with high-dimensional or non-linear datasets. Additionally, while white-box models strive for transparency, the level of human understanding can still be influenced by the complexity of the problem and the individual's expertise.

*2) The Motivation of Explaining Black-Box Models*

While black-box models can achieve high predictive accuracy, their lack of transparency poses significant challenges and risks. The first to think about building trust and acceptance, building trust and acceptance in machine learning models is paramount, especially when dealing with black-box models that often appear opaque and untrustworthy. Users are more inclined to trust and accept a model's decisions if they comprehend the underlying reasoning. Without clear explanations, these models can be perceived as unreliable and unpredictable, ultimately hindering their adoption and limiting their practical value. Providing transparent and understandable explanations bridges this trust gap, empowering users to confidently embrace and utilize the model's predictions[29].

The next consideration, ensuring fairness and mitigating bias, is a critical aspect of responsible machine learning, particularly with black-box models. These models can inadvertently perpetuate and even amplify biases present in the training data, leading to unfair or discriminatory outcomes. Interpreting these models allows us to shine a light on these potential biases, understand how they influence predictions, and take steps to address them. By promoting transparency and understanding, we can strive for fairer, more ethical decision-making in machine learning applications[8]. Then, fostering user confidence and encouraging wider adoption of AI systems hinges on transparency and understanding. When users can grasp how a model operates and arrives at its conclusions, they are more likely to trust its recommendations and utilize it effectively. Explainability plays a crucial role in this process, bridging the gap between complex algorithms and human comprehension. By providing clear and accessible explanations, we empower users to engage with AI systems confidently, leading to greater acceptance and integration of these powerful technologies[24].

When a black-box model makes an error, it can be challenging to pinpoint the root cause without a clear understanding of its internal decision-making process. Explainability in machine learning provides the necessary transparency to identify the source of these errors, allowing developers to make targeted improvements and build more robust systems[30]. Furthermore, industries like healthcare and finance operate under strict regulatory frameworks that demand transparency and accountability in decision-making. Explaining black-box models is essential for demonstrating compliance with these regulations and ensuring that decisions are made fairly and ethically[5].

Explaining black-box models transcends mere curiosity; it's about fostering trust with users, guaranteeing fairness in decision-making, enhancing model accuracy, and adhering to ethical and regulatory guidelines. By prioritizing transparency and interpretability, we can fully leverage the power of these advanced tools while effectively mitigating potential risks.

*C. Quantitative Metrics for Explainability*

When assessing the interpretability or explainability of phishing detection models through a comparative analysis of black-box and white-box approaches, it is crucial to consider specific metrics that capture the extent to which the model's predictions can be comprehended by humans. The following interpretability/explainability metrics can be particularly valuable in this context:

1. Fidelity: This metric evaluates the degree to which an interpretability approach approximates the predictions of the original model. High fidelity indicates that the explanation accurately represents the model's behavior. For instance, if employing a surrogate model to explain a black-box, the surrogate's predictions should closely align with those of the black-box.[31]

2. Simplicity: This metric reflects the complexity of the explanation provided. Generally, a simpler explanation is more interpretable for users. In the context of phishing detection, this can entail using a smaller number of rules or conditions to explicate why an email is classified as phishing or benign[32].

3. Comprehensiveness: This metric evaluates the degree to which the explanation encompasses the salient aspects of the input data or model components essential for the prediction. For example, in the context of phishing detection, a comprehensive explanation should cover the key indicators, such as URL irregularities or suspicious textual patterns, that contribute to the classification decision[11].

4. Consistency: This metric assesses whether the model provides consistent explanations for similar input instances. Maintaining consistency in the interpretability of the phishing detection system is crucial, as it ensures that analogous phishing indicators result in analogous explanations across different cases [33].

5. Accuracy of Explanations: When employing post-hoc explanation techniques, such as LIME or SHAP, the accuracy of the explanations evaluates the degree to which the interpretable model's output corresponds to the primary model's decisions[34].

6. Stability: This metric evaluates the consistency of the explanations provided by the model, ensuring that minor variations in the input data do not result in significantly different explanations. In the context of phishing detection, this is a crucial factor to consider, as it helps maintain the reliability and trustworthiness of the system by providing consistent insights, even when faced with slight changes in the email content[19].

7. Actionability: This metric evaluates the degree to which the explanations offered by the model can inform tangible, actionable steps for users. In the context of phishing detection, this may entail providing insights on the specific features or indicators that signal the presence of phishing, thereby empowering security teams to undertake proactive preventive measures[35].

Selecting and weighing these metrics based on the model, audience's expertise, and desired transparency will ensure the explanations are informative and useful for phishing detection. These metrics offer a structured way to assess and compare the explainability of different machine learning models, helping developers choose the best approach for their specific needs and users.

### III. METHODS

#### A. Analytical Review of Black-Box and White-Box Comparison

The approach of a comparative analysis of black-box and white-box models to evaluate their suitability for explainable phishing detection. For black-box models, we analyze common algorithms such as Deep Neural Networks (DNNs), Random Forests, Gradient Boosting Models (e.g., XGBoost), Support Vector Machines (SVMs) with non-linear kernels, and ensemble methods like stacked models. Each model is assessed based on interpretability metrics, including fidelity, simplicity, comprehensiveness, consistency, accuracy of explanations, stability, human interpretability, and actionability. Additionally, suitable Explainable AI (XAI) techniques, such as LIME, SHAP, Grad-CAM, Integrated Gradients, Tree SHAP, Permutation Feature Importance, Partial Dependence Plots (PDPs), and Individual Conditional Expectation (ICE) plots, are mapped to the respective models to enhance transparency and provide actionable insights.

For white-box models, the study evaluates Decision Trees, Logistic Regression, Rule-Based Models, k-Nearest Neighbors (k-NN), and Linear SVMs. The models are analyzed against the same interpretability metrics to understand their inherent strengths in explainable phishing detection. Decision Trees and Rule-Based Models, known for their high human interpretability and actionability, are particularly emphasized for their transparency and ability to generate clear, actionable rules. Logistic Regression and Linear SVMs are noted for their consistency and stability, though they may be less capable of capturing the complexities inherent in phishing detection. In contrast, k-NN, while consistent, lacks explicit rules, limiting its interpretability.

By systematically mapping the metrics and applying appropriate XAI techniques, the study bridges the gap between model performance and human understanding, providing a foundation for trust and actionable decision-making in high-stakes phishing detection scenarios.

#### B. Experimental Validation

Validation process by experiment shall carry out to strengthen the comparative analysis, to fulfill the process, it should be designed as follows:

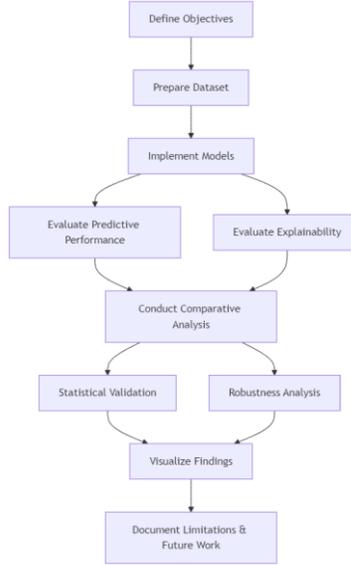

Fig. 1 Methodology

This validation process aims to compare the predictive accuracy and explainability of black-box and white-box models for phishing detection. The process begins by defining clear objectives for the comparison, focusing on both predictive power and interpretability. A phishing dataset will be meticulously prepared, ensuring data cleanliness and appropriate partitioning into training, validation, and test sets. Both black-box models (e.g., XGBoost) and white-box models (e.g., Explainable Boosting Machine) will be trained using consistent methodologies and optimized hyperparameters. Predictive performance will be rigorously evaluated using metrics such as accuracy, precision, recall, False Positive rate, and AUC-ROC on the held-out test set. Explainability will be assessed using XAI techniques like SHAP and LIME for black-box models, while the inherent interpretability of white-box models will be directly analyzed using metrics like fidelity, simplicity, and comprehensiveness. A comparative analysis will then be conducted, considering quantitative performance metrics. Statistical tests will validate the significance of observed differences between models. Robustness analysis will be performed to ensure the stability and generalizability of both model predictions and explanations using adversarial examples, unseen data, or noisy inputs. The findings will be visualized using charts and diagrams for explainability, and chats for performance comparisons. Finally, the study will summarize the key findings, acknowledge limitations, and propose future research directions, such as evaluating performance on diverse datasets or exploring hybrid model approaches.

## IV. RESULTS AND FINDINGS

### A. Black-Box and White-Box Model Comparative Analysis

In high-stakes fields like phishing detection, understanding the "how" behind a black-box model's decisions is critical. Let's compare common black-box models, focusing on metrics that gauge their interpretability and highlighting suitable XAI techniques to make their inner workings more transparent from various sources [36], [37], [38], [39], [40].

TABLE 1
BLACK-BOX MODEL METRIC ASSESMENT

| Metric / XAI Method | Deep Neural Networks | Random Forests | Gradient Boosting Models (e.g., XGBoost) | Support Vector Machine (Non-linear) | Ensemble Methods (e.g., Stacked Models) |
|---|---|---|---|---|---|
| **Fidelity** | Moderate | Moderate | High | Moderate | Moderate |
| **Simplicity** | Low | Low | Low to Moderate | Low | Low |
| **Comprehensiveness** | High | High | High | Moderate | High |
| **Consistency** | Moderate | Moderate | Low to Moderate | High | Low to Moderate |
| **Accuracy of Explanations** | High | Moderate | High | High | High |
| **Stability** | Moderate | Moderate | Moderate | High | Low to Moderate |
| **Human Interpretability** | Low | Moderate | Moderate | Low | Low |
| **Actionability** | Low | Moderate | Moderate | Low | Low |
| **Suitable XAI Methods** | LIME, SHAP, Grad-CAM, Integrated Gradients | LIME, SHAP, TreeSHAP | SHAP, LIME, Partial Dependence Plot (PDP) | LIME, SHAP, Local Interpretable Projections | SHAP, LIME, Permutation Feature Importance |

Deep Neural Networks, while powerful, often act as "black boxes" in phishing detection, achieving high fidelity but lacking transparency. However, XAI techniques like SHAP, LIME, Integrated Gradients, and LRP can shed light on their decision-making processes. Similarly, ensemble methods like Random Forests and Gradient Boosting Machines, known for their accuracy, benefit from XAI methods like Tree SHAP, Permutation Feature Importance, PDPs, and ICE plots to enhance interpretability. Support Vector Machines with non-linear kernels, while effective, can be understood better using local explanations from LIME and SHAP, along with counterfactual examples. In essence, applying the right XAI methods to these complex models bridges the gap between powerful predictions and human understanding, crucial for building trust and enabling effective action in phishing detection.

Mapping the quantitative metrics to white-box models helps us understand which algorithms are best suited for explainability phishing detection, based on each model's inherent strengths. Here's how some well-known white-box algorithms perform with respect to explainability metrics which resume from [41], [42], [43]:

TABLE 2
WHITE BOX MODEL METRIC ASSESSMENT

| Metric | Decision Trees | Logistic Regression | Rule-Based Models | k-Nearest Neighbors (k-NN) | Explainable Boosting Machine |
|---|---|---|---|---|---|
| **Fidelity** | High | High | High | Moderate | High |
| **Simplicity** | High (for shallow trees) | Moderate to High | High | Moderate to Low | Moderate |
| **Comprehensiveness** | Moderate to High | Limited | High | High | High |
| **Consistency** | Moderate | High | Moderate | High | High |
| **Accuracy of Explanations** | High | High | Moderate to High | Moderate | High |
| **Stability** | Low to Moderate | High | Moderate | High | High |
| **Human Interpretability** | High | High | High | Moderate | High |
| **Actionability** | High | Moderate | High | Low | High |

Key observations highlight the strengths of different models for interpretable phishing detection. Decision Trees, with their inherent transparency, especially when kept shallow, and Rule-Based Models, offering clear and actionable rules, emerge as particularly advantageous. Their high human interpretability and actionability make them well-suited for phishing detection, where understanding the reasoning behind classifications is paramount. While Logistic Regression provides consistent explanations and Linear SVMs excel in linearly separable data, they may not fully capture the complexities of phishing detection. K-Nearest Neighbors, though consistent, lack explicit rules, hindering interpretability.

*B. Experimental Result*

Based on several dataset that taken from Kaggle.com and data.mendeley.com, this work result as follows:
1. XGBoost

Table 1 XGBoost Performance Result

| Dataset Name | # Instances | # Features | Accuracy | Precision | Recall | FP Rate | ROC AUC | Runtime (seconds) |
|---|---|---|---|---|---|---|---|---|
| **ds_100K20** | 100077 | 20 | 89,68% | 88,15% | 91,60% | 12,23% | 96,32% | 296,88 |
| **ds_10K18** | 10000 | 18 | 99,85% | 100,00% | 99,70% | 0,00% | 100,00% | 94,15 |
| **ds_10K50** | 10000 | 49 | 98,95% | 98,63% | 99,31% | 1,42% | 99,91% | 78,46 |
| **ds_11055_rev** | 11055 | 32 | 97,16% | 96,85% | 97,62% | 3,32% | 99,68% | 1,94 |
| **ds_11055** | 11055 | 31 | 97,44% | 97,24% | 97,78% | 2,91% | 99,67% | 1,63 |
| **ds_11K89** | 11481 | 89 | 98,65% | 98,66% | 98,58% | 1,28% | 99,71% | 2,50 |
| **ds_129K112** | 129698 | 112 | 97,45% | 97,26% | 97,65% | 2,76% | 99,68% | 21,93 |
| **ds_235795_54** | 235795 | 55 | 99,99% | 99,99% | 100,00% | 0,01% | 100,00% | 28,83 |
| **ds_247950** | 247950 | 42 | 90,99% | 93,23% | 88,37% | 6,39% | 96,91% | 23,32 |
| **ds_600K11** | 662591 | 10 | 82,55% | 80,63% | 85,73% | 20,65% | 90,80% | 9,16 |
| **ds_88K112** | 88647 | 112 | 97,59% | 97,36% | 97,79% | 2,59% | 99,67% | 4,23 |
| **ds_90K32** | 90000 | 34 | 99,99% | 100,00% | 99,99% | 0,00% | 100,00% | 0,78 |

2. Explainable Boosting Machine

Table 2 Explainable Boosting Machine Performance Result

| Dataset Name | # Instances | # Features | Accuracy | Precision | Recall | FP Rate | ROC AUC | Runtime (seconds) |
|---|---|---|---|---|---|---|---|---|
| **ds_100K20** | 100077 | 20 | 88,97% | 87,60% | 90,69% | 12,75% | 95,99% | 256,98 |
| **ds_10K18** | 10000 | 18 | 99,95% | 100,00% | 99,90% | 0,00% | 100,00% | 4,58 |
| **ds_10K50** | 10000 | 49 | 98,30% | 97,85% | 98,81% | 2,23% | 99,81% | 12,14 |
| **ds_11055_rev** | 11055 | 32 | 95,78% | 95,44% | 96,35% | 4,82% | 99,40% | 9,89 |
| **ds_11055** | 11055 | 31 | 94,76% | 94,14% | 95,71% | 6,23% | 99,11% | 12,01 |
| **ds_11K89** | 11481 | 89 | 98,00% | 97,87% | 98,04% | 2,04% | 99,55% | 26,58 |
| **ds_129K112** | 129698 | 112 | 97,65% | 97,73% | 97,58% | 2,28% | 99,67% | 3.581,70 |
| **ds_235795_54** | 235795 | 55 | 99,99% | 99,99% | 99,99% | 7,40% | 100,00% | 314,90 |
| **ds_247950** | 247950 | 42 | 89,22% | 91,35% | 86,62% | 8,18% | 95,81% | 2.342,93 |
| **ds_600K11** | 662591 | 10 | 79,68% | 77,44% | 83,82% | 24,48% | 87,60% | 10.804,34 |
| **ds_88K112** | 88647 | 112 | 97,20% | 97,08% | 97,27% | 2,87% | 99,56% | 453,78 |
| **ds_90K32** | 90000 | 34 | 99,99% | 100,00% | 99,99% | 0,00% | 100,00% | 44,08 |

Based on result above, mapping to metric that explain in previous part, here is the result:

Table 3 EBM and XGBoost Performance Metric Result

| Metric | EBM | XGBoost |
|---|---|---|
| **Fidelity** | High (except Moderate for ds_600K11_rev.csv) | High (except Moderate for ds_600K11_rev.csv) |
| **Simplicity** | Moderate (reduced for large datasets) | Moderate (due to XGBoost's complexity) |

| Metric | EBM | XGBoost |
|---|---|---|
| **Comprehensiveness** | High (except Moderate for ds_600K11_rev.csv) | High (except Moderate for ds_600K11_rev.csv) |
| **Consistency** | High (with variability in challenging datasets) | High (with slight variability in challenging cases) |
| **Stability** | High (except Moderate for noisy datasets) | High (except Moderate for noisy datasets) |

The other metric explained by SHAP method and interpreted from this guidance:
  a) **Stability**: Analyze the spread of SHAP values for each feature. Narrow spreads indicate high stability; wide spreads indicate low stability.
  b) **Consistency**: Check for patterns in SHAP values relative to feature values. Consistent features show predictable SHAP value behavior.
  c) **Accuracy of Explanation:** Validate the feature ranking and color gradients against domain knowledge. Do high-impact features match expectations?
  d) **Interpretability:** Evaluate the clarity of feature ranking and the distinction in SHAP value patterns (e.g., red = high value, blue = low value).
  e) **Actionability**: Identify top-ranked features with clear trends (e.g., red values causing positive SHAP contributions suggest specific actions).

The analysis represented by dataset which have the most instance numbers but less features such ds_600K11, high instances and high feature numbers such as ds_129K112, less instances and less feature such as ds_10K18 and less instances and more feature such as ds_11K89.

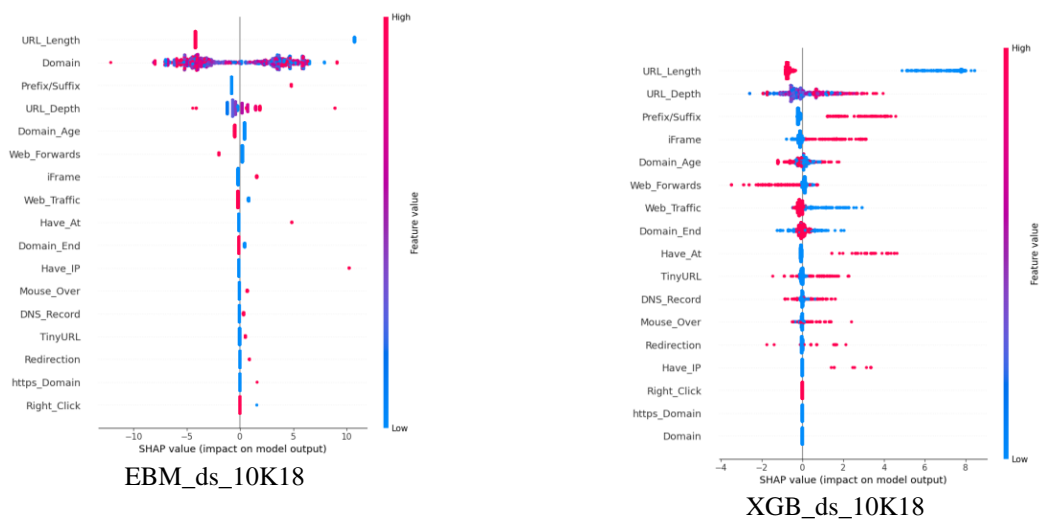

EBM_ds_10K18          XGB_ds_10K18

Fig. 2 EDM and XGB in ds_11055 Explanation Plot

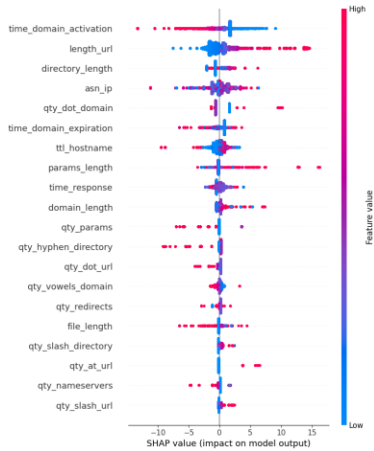 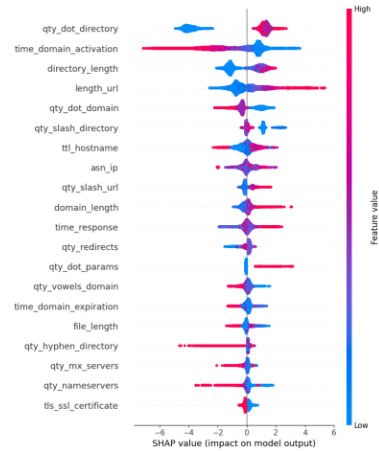

EBM_ds_129K112          XGB_ds_129K112

Fig. 3 EBM and XGB in ds_129K112 Explanation Plot

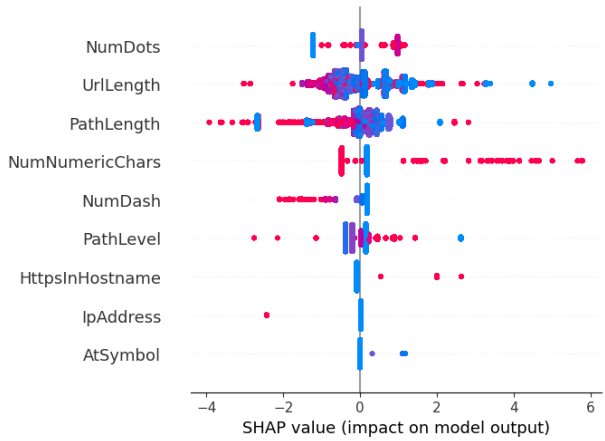 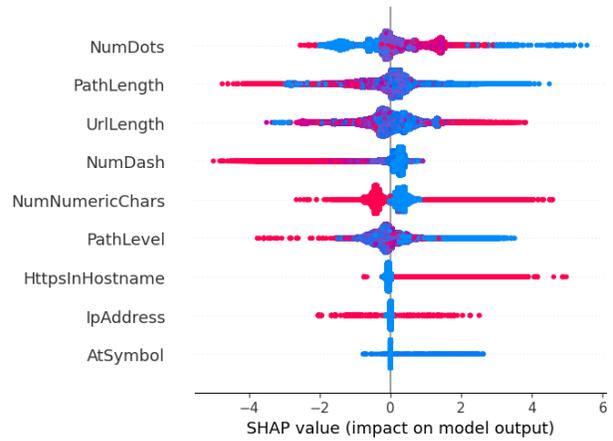

EBM_ds_600K11          XGB_ds_600K11

Fig. 4 EBM and XGB in ds_600K11 Explanation Plot

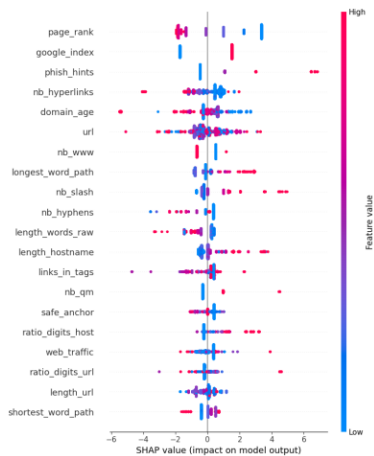 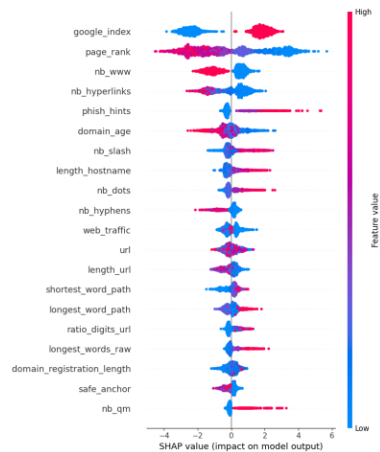

EBM_ds_11K89          XGB_ds_11K89

Fig. 5 EBM and XGB in ds_600K11 Explanation Plot

| Metric | EBM | | XGBoost | |
|---|---|---|---|---|
| | Explanation | Rating | Explanation | Rating |
| **Stability** | Overall dataset has better stability than XGBoost | Moderate to High | . Overall datasets have wider spread of SHAP value distribution that mean stability outperformed that EBM | Moderate |
| **Consistency** | Overall datasets show low to moderate consistency since the distribution SHAP value are diverse both negative and positive impact | Low to Moderate | Overall datasets show low to moderate consistency since the distribution SHAP value are diverse both negative and positive impact | Low to Moderate |
| **Accuracy of Explanation** | Several datasets has shown high accuracy explanation in most features of ds_10K18 except domain and URL_Dept. Same high accuracy shown in ds_11K89 for page_rank, google_index and phis_hint | Low to Moderate | Overall, most datasets has shown low accuracy explanation | Low |
| **Interpretability** | Most features has more clarity to interpreted in ds_10K18 and 600K11but the other dataset has lower clarity to interpreted | Low to Moderate | .The clarity to interpreted for more features shown at ds_10K18 and 600K11, the other features hard to interpreted. | Low to Moderate |
| **Actionability** | The most feature in ds_10K18 have better actionability than XGB. The other dataset. In ds_11K89, page_rank, google_index, nb_www has better actionability than XGB. | Moderate | Overall, most feature in datasets have less actionability than EBM | Low to Moderate |

## V. DISCUSSION

Important Points to Note of fidelity metric, with the exception of ds_600K11_rev.csv, where both models struggle presumably because of the dataset's properties, both models exhibit remarkably high performance across datasets. Then, in simplicity metric, both are quite easy; XGBoost's boosting introduces intrinsic complexity, whereas EBM is easier to understand but becomes less straightforward with larger datasets. In terms of comprehensiveness, except for edge situations in datasets such as ds_600K11_rev.csv, both models fully explain patterns. Other metric e.g consistency, in difficult datasets, both show a high degree of consistency with some fluctuation. Finally, stability metric, both are stable across datasets, while noisy data may cause them to falter. Here is the conclusion regarding performance metric dimensions, EBM and XGBoost have very similar fidelity, comprehensiveness, consistency, and stability.

In explainability metric perspective, for stability metric, EBM does better than XGBoost because its SHAP value distributions are smaller, which means that its feature contributions are more stable. Then consistency metric, the SHAP value distributions for most features are different, so both models have low to middling consistency. In explanation accuracy, EBM is better at explaining how features affect certain datasets, while XGBoost has trouble giving acceptable reasons. For interpretability metric, it's about the same for both models to understand features in datasets like ds_10K18. However, it's harder for both models to understand features in other datasets.Finally, actionability: EBM gives more useful information that can be used, especially in datasets like ds_10K18 and

ds_11K89, while XGBoost gives less useful information in general. The conclusion are, EBM has a slight interpretability edge over XGBoost due to its naturally interpretable structure, while XGBoost forgoes simplicity in lieu of increased scalability and versatility. EBM outperforms XGBoost in terms of actionability, explanation accuracy, and stability. Although there is room for improvement in both models over a range of datasets, their consistency and interpretability are comparable. In general, applications needing explainability and actionable insights are better suited for EBM.

To deploy interpretable and trustworthy machine learning models for phishing detection effectively, developers should consider the following practical implications and approach:

1. Iterative Model Refinement: Continuously monitor the performance and user feedback of the deployed phishing detection system. Regularly update the model and explanations to address any issues or changing user requirements, ensuring the system remains reliable and trusted over time[39][40].

2. Contextual Explanations: Tailor the explanations to the specific needs and technical expertise of the target users. For example, provide more detailed technical explanations for security professionals, while offering simpler, more user-friendly explanations for general end-users[41][42].

3. Integrated Workflow: Seamlessly integrate the interpretable phishing detection system into the existing workflows and tools used by end-users. This will help to ensure a smooth adoption and enhance the overall user experience.[43]

4. Collaborative Development: Engage with end-users throughout the development process to gather feedback and incorporate their insights. This collaborative approach will help to ensure the explanations are meaningful and useful for the target audience[44].

5. Ongoing Training and Support: Provide comprehensive training and support materials to help end-users understand the capabilities and limitations of the interpretable phishing detection system. This will empower users to make informed decisions and build trust in the model's recommendations[24][45].

By following this practical approach, developers can successfully deploy interpretable and trustworthy machine learning models for phishing detection, fostering user confidence and ensuring the effective implementation of these advanced technologies.

## VI. CONCLUSIONS

EBM and XGBoost work about the same on all datasets when it comes to accuracy, simplicity, coverage, uniformity, and stability. Because its structure is naturally interpretable, EBM is a little easier to understand than XGBoost. On the other hand, XGBoost gives up simplicity for more scalability and freedom. EBM is better than XGBoost when it comes to being stable, actionable, and accurate. Both models are comparable in terms of interpretability and consistency, with room for improvement in diverse datasets. **EBM** is generally better suited for applications requiring explainability and actionable insights. Finally, each model, white-box and black-box model has positive and negative aspects both for performance metric and for explainable metric. It is important to consider the objective of model usage.

For developers to successfully use machine learning models that can be trusted to spot phishing, they should think about the following practical implications:

1. Iterative Model Refinement: Keep an eye on the phishing detection system's performance and user comments all the time to make sure it stays reliable and trustworthy.
2. Contextual Explanations: Make sure the explanations fit the needs and level of professional knowledge of the people who will be using them.
3. Integrated Workflow: The interpretable phishing detection system can be easily added to end users' existing tools and processes.
4. Collaborative Development: Talk to end users during the whole development process to get their comments and use their ideas.
5. Ongoing Training and Support: Give end users thorough training and support tools to help them understand what the interpretable phishing detection system can and can't do.